\documentclass{iopconfser}

\usepackage{url}
\usepackage[symbol]{footmisc}
\usepackage[english]{babel}       
\usepackage[utf8]{inputenx}     

\usepackage{color}                
\usepackage{graphicx}             
\usepackage{siunitx}
\usepackage{enumitem}			
\usepackage{float}              
\usepackage{subfigure}          
\usepackage[skip=2pt]{caption}	
\usepackage{subfigure}           
\usepackage{amsmath}            

\usepackage[square, numbers, sort&compress]{natbib}				
\usepackage{doi}				

\usepackage{hyperref}                            
\definecolor{LinkColor}{rgb}{0,0,0.75}           
\hypersetup{
	pdftitle = {AtmoHEAD 2024 Proceeding of Weitz, Melanie Joan},
	pdfsubject = {AtmoHEAD 2024 Proceeding of Weitz, Melanie Joan},
	pdfkeywords = {AtmoHEAD 2024, Proceeding, TGFs, Uni Wuppertal},
	pdfauthor = {Melanie Joan Weitz for the Pierre Auger Collaboration},
	bookmarksnumbered = true,
	bookmarksopen = false,
	colorlinks = false,
	hypertexnames = true,
	a4paper = true,
	colorlinks=true,%
	linkcolor=LinkColor,%
	citecolor=LinkColor,%
	filecolor=LinkColor,%
	menucolor=LinkColor,%
	pagecolor=LinkColor,%
	urlcolor=LinkColor,
	breaklinks=true	
}

\usepackage[nameinlink, noabbrev]{cleveref} 

\usepackage{ragged2e}

\begin{document}

\title{Adding interferometric lightning detection to the\\
Pierre Auger Observatory}

\author{\renewcommand{\thefootnote}{\fnsymbol{footnote}}Melanie Joan Weitz\footnote[1]{Speaker}$^{,a}$ for the Pierre Auger Collaboration$^b$}

\affil{$^a$Bergische Universität Wuppertal, Wuppertal, Germany}
\affil{$^b$Observatorio Pierre Auger, Av. San Martín Norte 304, 5613 Malargüe, Argentina}

\email{spokespersons@auger.org}
\fullauthorlist{\url{https://www.auger.org/archive/authors_2024_10.html}}

\begin{abstract}
\justifying
The Pierre Auger Observatory has detected downward terrestrial gamma-ray flashes (TGFs) with its Surface Detector. A key to understanding this high-energy radiation in thunderstorms is to combine such measurements with measurements of lightning processes in their earliest stages. With eleven modified Auger Engineering Radio Array (AERA) stations we can build an interferometric lightning detection array working in the bandwidth between 30 – \qty{80}{\mega\hertz} inside the Surface Detector array to precisely measure lightning stepped leaders in 3D. These measurements allow us to decipher the cause of TGFs and clarify the reason for the observed high-energy particles in thunderstorms.
We will present the current status of the detection plans including the configuration of the interferometric lightning detection array and the steps to take as well as the reconstruction characteristics obtained with AERA.
\end{abstract}

\renewcommand{\thefootnote}{\arabic{footnote}}
\section{The Pierre Auger Observatory and Terrestrial Gamma-ray Flashes}

\justifying
\setlength{\parindent}{20pt}
The Pierre Auger Observatory \cite{PAO} is the world's largest cosmic ray observatory with a detector area of \qty{3000}{\kilo\metre^2} located in the Province of Mendoza, Argentina. With its hybrid detection array it is designed to study the origin and characteristics of ultra-high energy cosmic rays with energies above \qty{e17}{\electronvolt}. Despite its focus the Pierre Auger Observatory offers large opportunities for the observation of high-energetic atmospheric phenomena like ELVES, halos \cite{MussaProceeding2023} as well as downward Terrestrial Gamma-ray Flashes (TGFs) \cite{ColalilloProceeding2023}. \par
TGFs are gamma-ray bursts produced by lightning from within the Earth's atmosphere and have a duration ranging from tens of microseconds up to milliseconds. After the first TGF observation from space via BATSE \cite{BATSE_TGF_Discovery} many other observations followed with first results of TGF characteristics \cite{DwyerReviewPaperHighEAtmosphericPhysics}. The TGF photons are produced via bremsstrahlung of energetic electrons interacting with air molecules. The production of this energetic electron avalanche was referred to the Relativistic Runaway Electron Avalanche (RREA) mechanism \cite{Babich1998}, where a single energetic seed electron initiates the production of secondary electrons which can slow down due to interactions with air molecules or can energize in a strong electric field of the thunderstorm. As soon as the rate of energy gain of electrons inside a thunderstorm cloud is higher than the rate of their energy loss due to ionization in air, they are able to ``runaway'' reaching relativistic energies and can produce additional energetic electrons resulting in an electron avalanche. However, neither the acceleration procedure of the seed electrons in the thundercloud, the involved lightning stage, nor its meteorological boundary conditions are clear in the TGF production mechanism. Two TGF production theories based on the RREA mechanism are under debate: the ``Lightning Leader" \cite{KoehnLightningLeader} and the ``Relativistic Feedback" models \cite{DwyerRelativisticFeedback}. Different studies regarding TGFs were made showing contradictory measurement results \cite{CelestinPasko2011, Xu2012, Dwyer2007, Dwyer2003}. \par
The Pierre Auger Observatory is contributing to these studies by the observation of downward TGFs with its water Cherenkov detectors showing results indicating with expectations of the ``Relativistic Feedback" model \cite{ColalilloProceeding2023}. The new insights can be broadened even further with the development of an Interferometric Lightning Detection Array consisting of eleven modified Auger Engineering Radio Array (AERA) stations in the area of the Pierre Auger Observatory. The bandwidth range from 30 to \qty{80}{\mega\hertz} can precisely measure lightning stepped leaders in 3D and could give insights into the connection between the earliest stages of lightning strikes, their meteorological boundary conditions and TGFs. The combination of the lightning stepped leader location and the direct location of the downward TGFs by the water Cherenkov detectors could give a first quantitative relation between a TGF source position and the triggering lightning.

\section{The Auger Engineering Radio Array and its first lightning analysis}\label{sec:AERAmeas}

\begin{figure}[htbp]
    \centering
    \includegraphics[width=.7\textwidth]{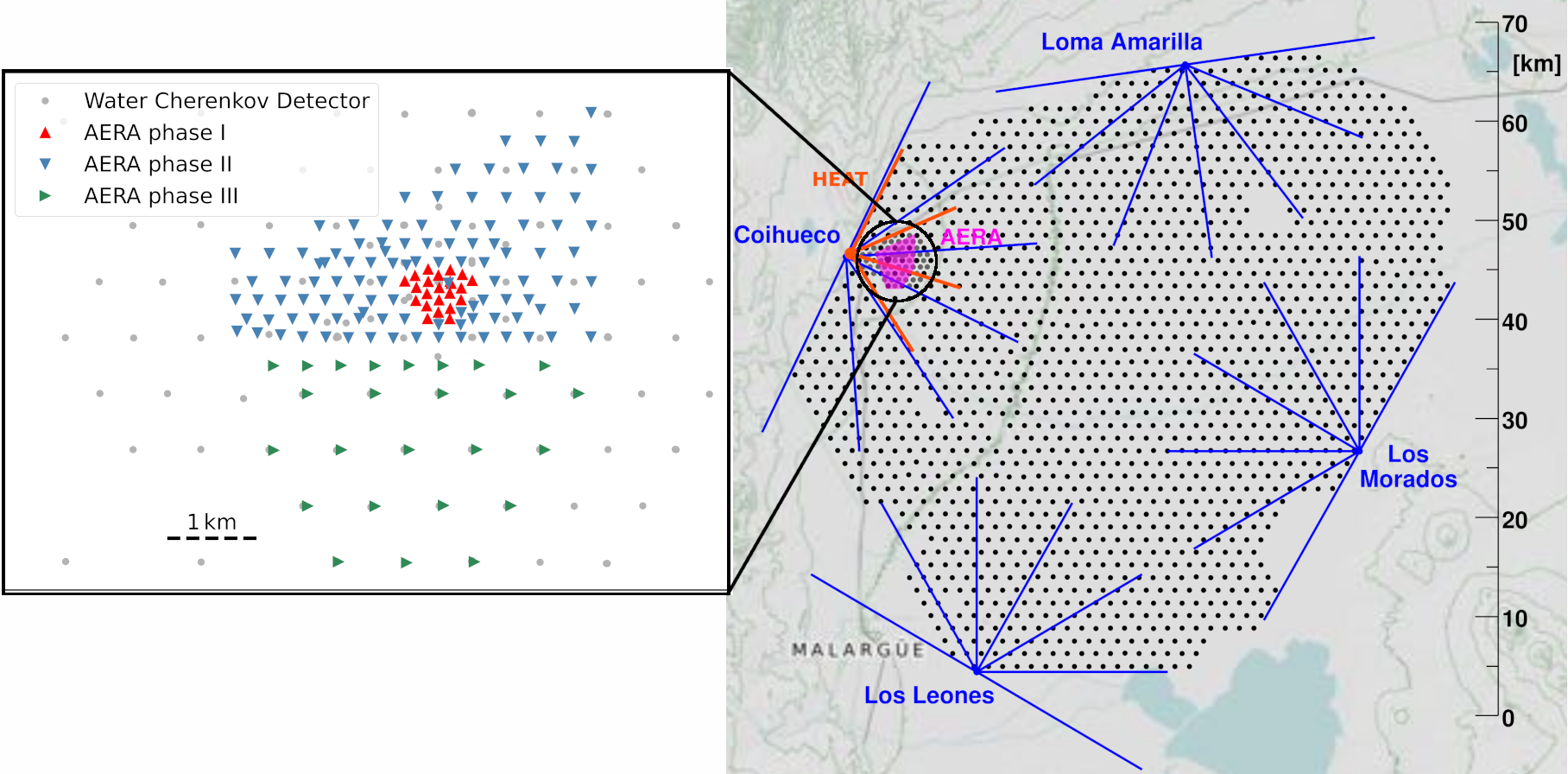}
    \caption{The Pierre Auger Observatory is shown at the right part of the figure together with a zoomed view of the AERA location on its left part. Right: Each black dot represents one of the 1660 water Cherenkov detectors in the Pierre Auger Observatory. The fluorescence telescopes at the four sites are color-coded in blue with their $\qty{30}{\degree}\,\text{x}\,\qty{30}{\degree}$ fields of view. Left: The AERA area is shown together with the spacing of the 154 radio detector stations. The light gray dots represent water Cherenkov detectors. The different color-coded triangles represent the AERA radio detector stations where the red color-coded ones show those of the AERA phase I, the green ones those from the AERA phase II and the blue ones those from the AERA phase III.}
    \label{fig:AERA_PAOmap}
\end{figure}

\begin{figure}[htbp]
    \centering
    \subfigure[LPDA]{\includegraphics[height=4.5cm]{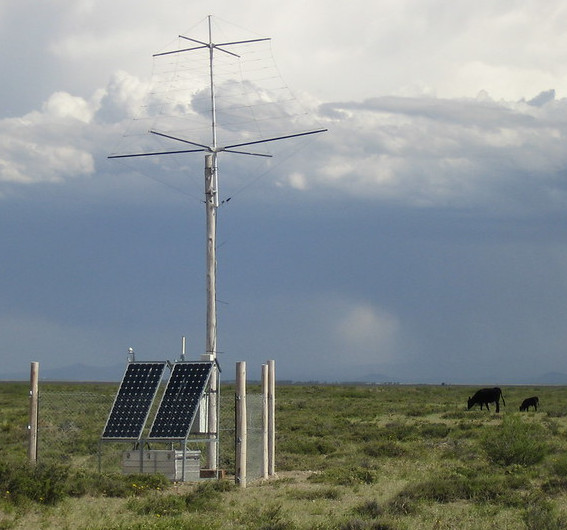}\label{fig:LPDA}}
    \subfigure[Butterfly]{\includegraphics[height=4.5cm]{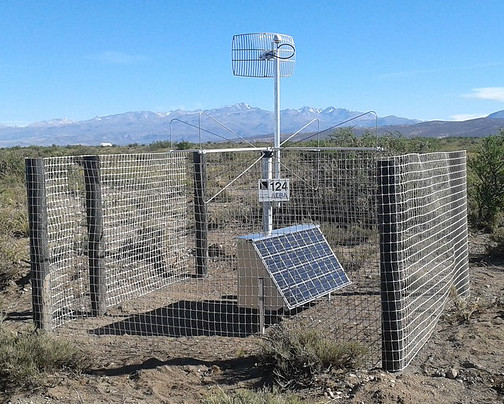}\label{fig:Butterfly}}
    \caption{The two different AERA station antenna types are shown. Left: Logarithmic Periodic Dipole Antenna (LPDA), right: Butterfly antenna.}
    \label{fig:AERA_antennatypes}
\end{figure}

The Auger Engineering Radio Array (AERA) measures short radio pulses of the order of nanoseconds emitted by cosmic ray air showers and is located near the Coihueco site of the Pierre Auger Observatory (cf. Figure \ref{fig:AERA_PAOmap}). AERA consists of 154 stations spaced over an area of approximately \qty{17}{\metre^2} and each station has a sensitive range of 30 to \qty{80}{\mega\hertz}. Two antenna types are used for the AERA stations: the Logarithmic Periodic Dipole Antenna (LPDA, cf. Fig. \ref{fig:LPDA}) and the Butterfly antenna (cf. Fig. \ref{fig:Butterfly}) \cite{Abreu2012}. Both antenna types are composed of two dipole antennas aligned in the geomagnetic North-South and East-West directions. The signal trace length is approximately \qty{11}{\micro\second} per dipole channel.\par
The AERA stations are promising candidates for lightning detection in the very high frequency range because of their sensitive range features.

Thunderstorms and their lightning are large disruption sources for the hybrid array of the Pierre Auger Observatory and its measurements. One of the disruption sources are the radio signals of the lightning impacting the measurements of the radio detectors. Lightning strikes emit radio signals in the low and the very high frequency band in the \unit{\kilo\hertz} and \unit{\mega\hertz} region, respectively. \par
\begin{figure}[htbp]
    \centering
    \subfigure[AERA lightning signal traces]{\includegraphics[width=.45\textwidth]{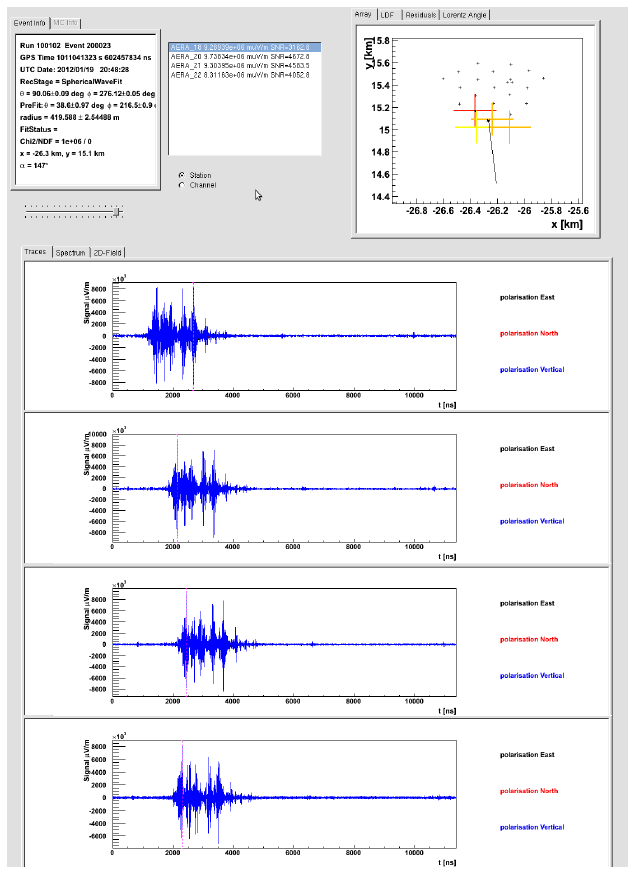}\label{fig:AERA}}
    \hspace{.2cm}
    \subfigure[Implemented Cross-correlation]{\includegraphics[width=.45\textwidth]{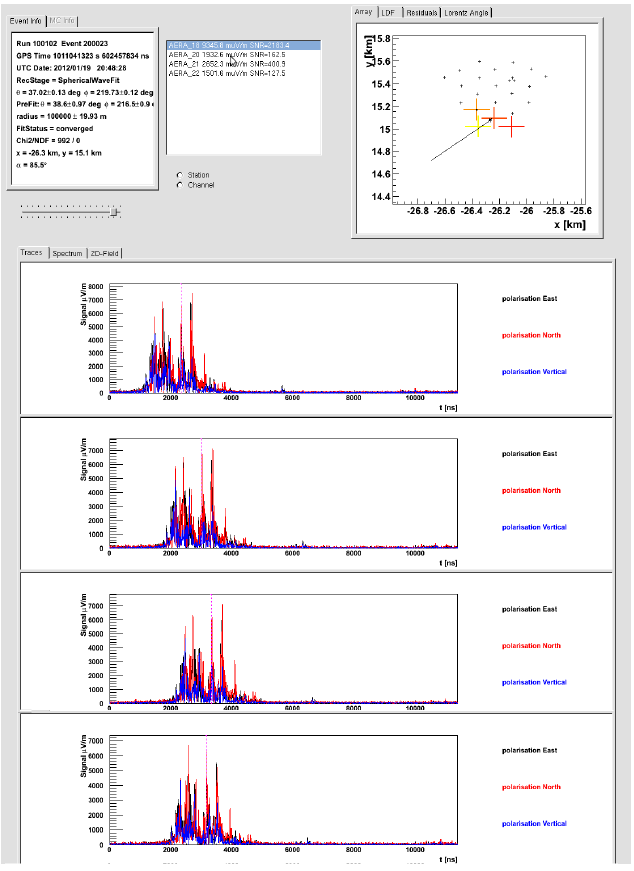}\label{fig:CrossCorrelation}}
    \caption{The reconstructed AERA event on January 19, 2012, is shown \cite{NiemietzDiss}. Left: The standard reconstruction in the Auger analysis framework and the lightning signal traces for different AERA stations are shown. One can see a bipolar signal for each station and the different time stamps indicated by the red dotted lines for each signal. Right: The cross-correlation method was implemented in the standard Auger analysis framework and was applied on the signal amplitude of the triggered AERA stations. One can see the same timestamp in the lightning signal ranges of each triggered AERA station based on their individual highest signal-product.}
    \label{fig:AERA_lightningNiemitz}
\end{figure}

A first analysis of lightning signals in the self-triggered AERA measurements was made by \cite{NiemietzDiss} during the AERA phase I development phase. Figure \ref{fig:AERA_lightningNiemitz} shows the analyzed lightning signal of January 19, 2012, in the standard Auger analysis framework with a time trace length of around \qty{11}{\micro\second}.\par
Figure \ref{fig:AERA} shows the bipolar lightning signal trace for each triggered AERA station. The red dotted line in each signal trace shows the result of the conventional analysis method for short radio pulses searching for the maximum amplitude to create a timestamp for each triggered station. This analysis method fails in case of lightning traces due to their long signal traces combined with multiple oscillations which can be seen in the different positions in the bipolar signals. As explained in \cite{RautenbergProceeding2015} one can use the cross-correlation method to find an optimal estimation of the timestamp for the reconstruction of the lightning events. The cross-correlation method was implemented in the standard Auger analysis framework and is shown in Figure \ref{fig:CrossCorrelation}. The cross-correlation method was applied to the amplitude of the signals of the triggered AERA stations and the optimal time stamp was found based on their highest signal-product. It can be seen in the position of the red dotted line which is at the same signal amplitude in the lightning signal ranges of each triggered AERA station.

This first analysis of self-triggered AERA measurements regarding lightning confirms that the basic idea of reusing AERA stations for lightning detection is working. The bandwidth range from 30 to \qty{80}{\mega\hertz} results in a resolution in the meter range allowing us to measure precisely lightning stepped leaders in 3D. 

\section{Planned configuration of the Interferometric Lightning Detection Array}

\begin{figure}[htbp]
    \centering
    \includegraphics[width=.7\textwidth]{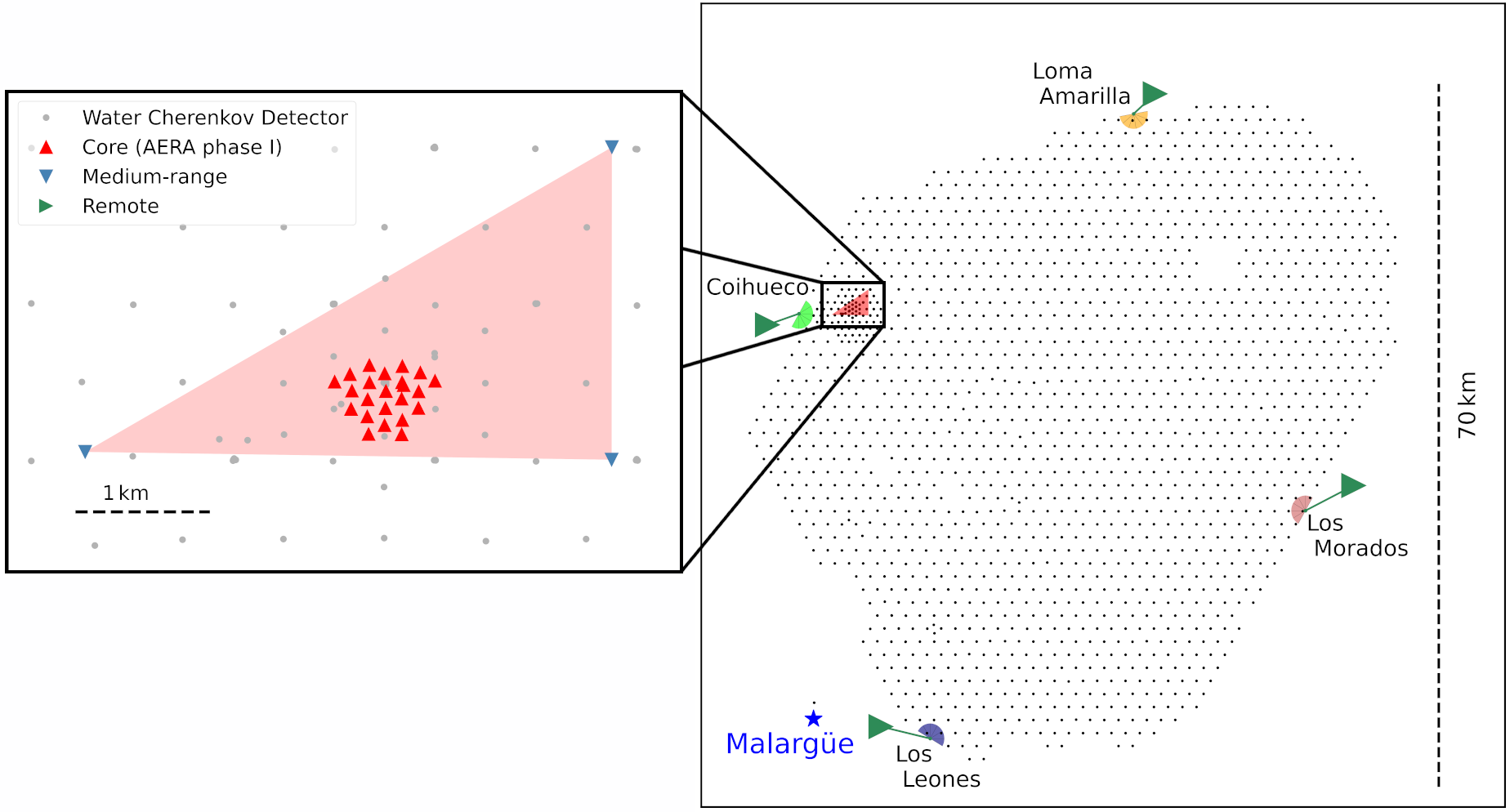} 
    \caption{The planned configuration of the Interferometric Lightning Detection Array is shown. The three different clusters are visualized with different color-coded triangles: the core in red, the medium-range in blue and the remote in green. The area of the medium-range cluster is filled light-red. Each light-gray circle represents a water Cherenkov detector.}
    \label{fig:ILDA_PAOmap}
\end{figure}

The planned configuration of the Interferometric Lightning Detection Array consists of three clusters with eleven modified AERA stations: the core, the medium-range and the remote cluster. A possible implementation at the Pierre Auger Observatory can be seen in Figure \ref{fig:ILDA_PAOmap} where each of the three clusters are visualized with different color-coded triangles.\par
The core cluster consists of four modified AERA stations with a baseline ranging from 58 to \qty{127}{\metre} and is color-coded in red. The core cluster is consistent with the area of AERA stations built during AERA phase I.\par
The medium-range cluster consists of three modified AERA stations with a baseline between 1.0 and \qty{2.5}{\kilo\metre} and is color-coded in blue. The outermost stations of AERA phase II were chosen for the visualization in Figure \ref{fig:ILDA_PAOmap} and their enclosed area is filled light-red.\par
The remote cluster consists of four modified AERA stations with baselines ranging from \qty{3.5}{\kilo\metre} up to \qty{66}{\kilo\metre} and is color-coded in green. Promising locations for the modified stations could be close to the four fluorescence telescope sites of the Pierre Auger Observatory.

\section{Future steps}\label{sec:StepsToTake}
The modification of the AERA stations is the first step to realize the Interferometric Lightning Detection Array. This includes the change of the time trace length from microseconds up to seconds for being able to measure a complete lightning event and the handling of the subsequent increasing amount of data. As a consequence, a new filter for the measurement trigger has to be developed. \par
Further investigations have to be made regarding the adjustment of the signal dynamical range to avoid possible saturation of the signal amplitude or its disappearance in noise. \par
The completion of the previously mentioned steps can lead to the first long trace read-out of a modified AERA station in November 2024.

\section{AERA Read-out Status}
A local test set-up is built including AERA boards for testing the current read-out status and for implementing the long signal trace. The read-out of the AERA boards seems to be flawless and the implementation of the long signal trace read-out is completed for up to \qty{1}{\second}.\par
One challenge is the slow read-out and the long dead time as its consequence. In addition, the sampling works faster than the read-out and can possibly overtake it. \par
As previously mentioned in Section \ref{sec:StepsToTake} the data handling is a very important challenge to solve. The size of one measurement trace is \qty{5.76}{\giga\byte} with a time trace length of \qty{8}{\second} and is responsible for the long read-out time together with the low communication bandwidth. As an example, one can estimate that a Wi-Fi bandwidth of \qty{22}{\mega\byte\second^{-1}} results in a read-out time of about \qty{5}{\minute}.

\section{Investigation of the Dynamical Range}
The adjustment of the signal dynamical range is one additional necessity to avoid the saturation of the signal or its disappearance in noise. An investigation with already existing AERA measurements has to be done to find lightning signal characteristics based on their signals for the review of a necessary adjustment of the station signal amplitude. \par
The outline for this investigation is the combination of the existing AERA measurements and an external lightning trigger. The external lightning trigger can be the reconstructed lightning events of the Lightning Detection System of Auger \cite{RautenbergProceeding2015} or the lightning flagged water Cherenkov detector data within a \qty{5}{\kilo\metre} distance of the AERA stations. Coincidences of the GPS timestamps between the external trigger and the AERA measurements could deliver lightning signal traces which can be investigated for characteristics.\par
The first attempt of the investigation of the signal dynamical range of the lightning AERA signal was made with the combination of self-triggered AERA measurements and the reconstructed lightning events by the Lightning Detection System of Auger. Since the first analysis described in Section \ref{sec:AERAmeas}, the building phase of AERA was finished and a new self-trigger algorithm was invented. However, this is the first analysis made with the new self-triggered AERA measurements regarding lightning. \par
The standard Auger analysis framework was modified to write out the self-triggered AERA signal traces and it was searched for coincidences in the GPS time. \par
The current challenge is the assignment of lightning traces in the AERA measurements because of few GPS time matches. One additional investigation has to be made if there is a systematic time offset.

\section{Summary}

Thunderstorms and lightning are important for the Pierre Auger Observatory because of their radio disruptions of the measurements made with water Cherenkov detector and radio detector stations. As a consequence the water Cherenkov detector lightning veto and the Lightning Detection System were developed. In addition the Pierre Auger Observatory offers large opportunities for studies of high-energetic atmospheric phenomena.\par
The first lightning mapping array was done with AERA but was not optimal because of its short trace length of approximately \qty{11}{\micro\second}.\par
The Interferometric Lightning Detection Array could be a promising attempt to enhance the understanding of thunderstorms and lightning and could give hints for their correlation with TGFs. 

\bibliography{bibliography}
\bibliographystyle{unsrt}

\end{document}